# Deterministic transfer of two-dimensional materials by all-dry viscoelastic stamping

*Andres Castellanos-Gomez \*, Michele Buscema, Rianda Molenaar, Vibhor Singh, Laurens Janssen,*

*Herre S. J. van der Zant and Gary A. Steele*

Kavli Institute of Nanoscience, Delft University of Technology, Lorentzweg 1, 2628 CJ Delft (The

Netherlands).

a.castellanosgomez@tudelft.nl

ABSTRACT

Deterministic transfer of two-dimensional crystals constitutes a crucial step towards the fabrication of heterostructures based on artificial stacking of two-dimensional materials. Moreover, control on the positioning of two-dimensional crystals facilitates their integration in complex devices, which enables the exploration of novel applications and the discovery of new phenomena in these materials. Up to date, deterministic transfer methods rely on the use of sacrificial polymer layers and wet chemistry to some extent. Here, we develop an all-dry transfer method that relies on viscoelastic stamps and does not employ any wet chemistry step. This is found very advantageous to freely suspend these materials as there are no capillary forces involved in the process. Moreover, the whole fabrication process is quick, efficient, clean, and it can be performed with high yield.

MANUSCRIPT TEXT

Since its isolation in 2004 [1], graphene has raised a huge interest in the scientific community [2]. This breakthrough also constitutes the origin of a new research topic devoted to the study of two dimensional atomic crystals [3]. One of the keys for the success of graphene research was the development of the micromechanical exfoliation method (so called Scotch tape method). Despite the simplicity of this method it can provide extremely high quality samples [4]. Nevertheless, the exfoliation method produces flakes with different sizes and thicknesses randomly distributed over the sample substrate, and only a small fraction of these flakes are atomically thin. The introduction of the optical identification method to find atomically thin crystals from the crowd of thicker bulky flakes constitutes the second key to guarantee the success of the graphene related research as it provides a fast, reliable and non-destructive way of locating the flakes [5-7]. The combination of mechanical exfoliation and optical identification, however, cannot provide a reliable way to fabricate more complex systems such as heterostructures formed by artificial stacking different 2D crystals [8-14]. The probability of creating these heterostructures by randomly depositing different 2D crystals onto the substrate is too small. To fabricate such systems new experimental approaches are needed to place the 2D crystals at a specific location [15, 16]. Three of these deterministic transfer methods are rather extended nowadays: the wedging method [17], the polyvinylalcohol (PVA) method [18] and the Evalcite method [19].

In the wedging transfer method water is used to lift off a hydrophobic polymer layer spin-coated onto a hydrophilic substrate. If the hydrophilic substrate was covered by flakes, they can be lifted off with the hydrophobic polymer layer and transferred to another substrate using water as the transfer-active component. After the transfer, the hydrophilic polymer layer has to be removed with solvents. In the PVA transfer method, the flakes are transferred onto a polymer sacrificial layer (spin-coated on a substrate previously treated with a water solvable PVA polymer layer). The substrate is floated on the surface of a deionized water bath and once the water-soluble polymer has dissolved, the substrate sinks to the bottom of the bath and the polymer layer is scooped and subsequently dried. The polymer layer is then mounted in a micromanipulator to transfer the flakes to an arbitrary substrate. After transfer, the polymer sacrificial layer is dissolved with solvents. Finally, in the Evalcite method, the flakes are transferred onto a glass slide which has been spin-coated with a low glass temperature polymer (Evalcite). The glass slide is mounted in a micromanipulator and the acceptor substrate is heated up to 75 ºC – 100 ºC. When the polymer touches the substrate it melts and adheres strongly to the surface, facilitating the transfer from the glass slide to the surface. After the transfer, the polymer layer has to be removed with solvents as in the previous methods.

All these transfer methods thus rely on sacrificial polymer layers and require wet process at some stage of the fabrication process. This characteristic may hamper their usability as the acceptor surface may contain structures sensitive to the chemicals employed or to the capillary forces involved in the process. For instance, fabrication of freely suspended structures by conventional wet transfer methods remains challenging as they tend to collapse by the capillary forces. An all-dry alternative transfer method is therefore necessary to widen the range of applications of the deterministic transfer methods and to avoid contamination of the fabricated structures. In this article we present an all-dry transfer method that relies on viscoelastic stamps. Two-dimensional crystals are transferred with this method without employing any wet chemistry steps. This is found very advantageous to freely suspend these materials as there are no capillary forces involved in the process. Fabrication of artificial heterostructres and freely suspended atomically thin layers and the transfer of 2D crystals onto arbitrary substrates are demonstrated.

The experimental setup employed to transfer two-dimensional crystals comprises an optical microscope (Olympus BX 51 supplemented with an Olympus DP25 digital camera) supplemented with large working distance optical objectives and a three axis micrometer stage to accurately position the stamp (see Figure 1(a)), very similar to the setups employed in other transfer methods [17-19]. We have also developed a dedicated setup to carry out the transfer process which shows even better performance than the modified optical microscope with a much lower price (see Supporting Information for more details of this dedicated transfer setup).

The stamp is a thin layer of commercially available viscoelastic material (Gelfilm from Gelpak) which is adhered to a glass slide to facilitate its handling. The two-dimensional crystals to be transferred are deposited onto the viscoelastic layer by mechanical exfoliation of the bulk layered crystal with Nitto tape (Figure 1(b)). The surface of the stamp is inspected under the optical microscope to select the thinner flakes due to their faint contrast under normal illumination. As the stamp is transparent, transmission mode can be used to determine the number of layers. Raman spectroscopy can also be carried out on the surface of the stamp to confirm the thickness of the flake [20, 21]. Once a thin flake has been identified, the acceptor substrate is fixed on the sample XYZ stage using double-sided tape. The stamp is then attached to the three axis manipulator with the flakes facing towards the sample. As the stamp is transparent, one can see the sample through it and thus it is possible to align the desired flake on the acceptor surface where one wants to transfer the flake with sub-micrometer resolution. A step-by-step guide of the transfer process can be found in the Supporting Information.

In order to transfer the flake to the acceptor surface, the stamp is pressed against the surface and it is peeled off very slowly. The working principle of the transfer is based on the viscoelasticity: the stamp behaves as an elastic solid at short timescales while it can slowly flow at long timescales [22]. Flakes are adhered to its surface because the viscoelastic material gets an intimate contact with the flakes. By slowly peeling off the stamp from the surface, the viscoelastic material detaches, releasing the flakes that adhere preferentially to the acceptor surface.

The potential of the presented technique is illustrated by transferring a few-layer graphene (FLG) flake onto h-BN. Figure 2(a) shows an example of a FLG flake deposited on the viscoelastic stamp. Figure 2(b) shows the h-BN flake, previously deposited on a Si/SiO$_2$ substrate by mechanical exfoliation (see materials and methods section in the Supporting Information), as seen through the viscoelastic stamp when the separation between the sample and the stamp is still large. When the stamp is brought closer and closer to the sample, the FLG flake appears more clearly as it gets more and more focused (Figure 2(c)). At this stage, it is still possible to align the flake to the h-BN flake using the XY knobs of the stamp stage. Once both flakes are aligned, the stamp is brought into contact with the acceptor surface, which can be clearly seen by a sudden change in color (Figure 2d). Once in contact, the stamp is not pressed further against the sample and it is peeled off very slowly (Figure 2e). Figure 2f shows the optical image of the transferred FLG flake after removing the stamp. The topography of the transferred flake is shown in Figure 2g, demonstrating that the FLG flake lays flat on the h-BN surface without bubbles or wrinkles. The whole transfer process can be accomplished in less than 15 minutes (see the Supporting Information for a real time video of a stamping process) [23].

The method can be applied to any kind of exfoliable layered crystals, allowing for almost infinite combinations of materials. We found that, with the dry transfer process, one can achieve a yield close to 100% when transferring onto atomically flat materials (to fabricate heterostructures, for instance). When rougher substrates are employed (such as samples with evaporated metals or with abrupt changes in their topography) the yield can be lower due to the reduced adhesion forces between the 2D material and the substrate. Figure 3 shows some examples of atomically thin heterostructures fabricated by stacking other two-dimensional materials employing this transfer technique. In Figure 3c, the transfer process has been repeated to fabricate a 'sandwiched' MoS$_2$ bilayer in between two h-BN flakes (see

Supp. Info.). Dashed lines have been used to highlight the different materials in the optical microscopy images (cartoons of the fabricated heterostructures are depicted above the optical images). The topography of the samples has been characterized by atomic force microscopy (AFM), which are shown above the optical microscopy images. Line profiles are included in the AFM images as insets to determine the thickness and the roughness of the transferred flakes. Note that although the fabricated heterostructures have not been subjected to any post-fabrication annealing step, their surfaces are clean of contaminants (typically present on samples fabricated by wet-transfer procedures).

We found that an excessive pressure applied during the transfer process may deform the viscoelastic stamp material yielding a high density of bubbles after the peeling off because of the sudden release of strain, similar to what has been observed in Ref. [24] (see Supporting Information). Nonetheless by carefully controlling the transfer pressure one can typically achieve large flat areas (as those shown in Figure 3) without wrinkles and bubbles, which is desirable for further fabrication of devices out of these heterostructures. The fabrication of heterostructures with large areas without bubbles and wrinkles can be achieved with yields of 30-40%. Section 9 of the Supporting Information presents the Raman spectroscopy characterization of some $MoS_2$ flakes transferred onto h-BN flakes. A thorough Raman spectroscopy and photoluminescence characterization of all-dry transferred heterostructures can be found in Ref. [25].

Another advantage of the all-dry transfer technique is that, due to the lack of capillary forces involved during the transfer procedure, fabricating freely suspended structures can be done straightforwardly. Figure 4a shows an example of a single-layer $MoS_2$ crystal that has been transferred onto a $SiO_2$/Si substrate pre-patterned with holes of different diameters. The single-layer $MoS_2$ is freely suspended over the holes forming micro-drumheads. The mechanical properties of these single-layer $MoS_2$ mechanical resonators fabricated by the stamping technique have been recently reported in Ref. [26].

The stamping method can be also applied to transfer two-dimensional crystals onto pre-fabricated devices with trenches and electrodes (see Figure 4b). Section 8 of the supporting information presents the electrical characterization of flakes transferred onto pre-designed circuits as well as vertical tunnel junctions fabricated by stacking different 2D materials. Our stamping transfer method has demonstrated to be very gentle, allowing the deposition of two-dimensional crystals even onto very fragile substrates. Figure 4c shows an example of a few-layer $MoS_2$ crystal transferred onto an AFM cantilever without damaging the cantilever. We have also successfully transfer two-dimensional materials onto silicon nitride membranes and holey carbon films, typically employed in transmission electron microscopy (see Section 7 of the Supporting Information).

In summary, we have introduced an all-dry transfer method that allows one to place two-dimensional crystals on a position desired by the user with sub-micron precision. As the process does not require any wet chemical steps, it can be used reliably to fabricate freely suspended structures and it significantly reduces the contamination in the fabricated samples. The potential of the proposed method has been illustrated by fabricating heterostructures formed by stacking different two-dimensional materials. The whole fabrication process can be accomplished in less than 15 minutes with a success yield close to 100%. Moreover, we find that about 30-40% of the fabricated heterostructures present large areas free of bubbles and wrinkles. The transfer method also offers the possibility to place two-dimensional crystals directly onto pre-fabricated circuit devices and micromechanical systems. Our all-dry transfer method has the potential to become a widely employed in nanotechnology, as it offers important advantages over alternative methods involving wet-chemical steps.

## ACKNOWLEDGMENT

A.C-G. acknowledges financial support through the FP7-Marie Curie Project PIEF-GA-2011-300802 ('STRENGTHNANO'). This work was supported by the European Union (FP7) through the program RODIN and the Dutch organization for Fundamental Research on Matter (FOM).

**Supporting Information Available**: Supporting Information includes: detailed explanation of the materials and methods employed, description of the dedicated deterministic transfer setup, real time video of the stamping process, fabrication of sandwiched structures, more examples of heterostructures, examples of samples with strain induced bubbles, Raman spectroscopy characterization of transferred flakes, electrical characterization of flakes transferred onto pre-patterned devices and examples of flakes transferred onto silicon nitride membranes and holey carbon films.

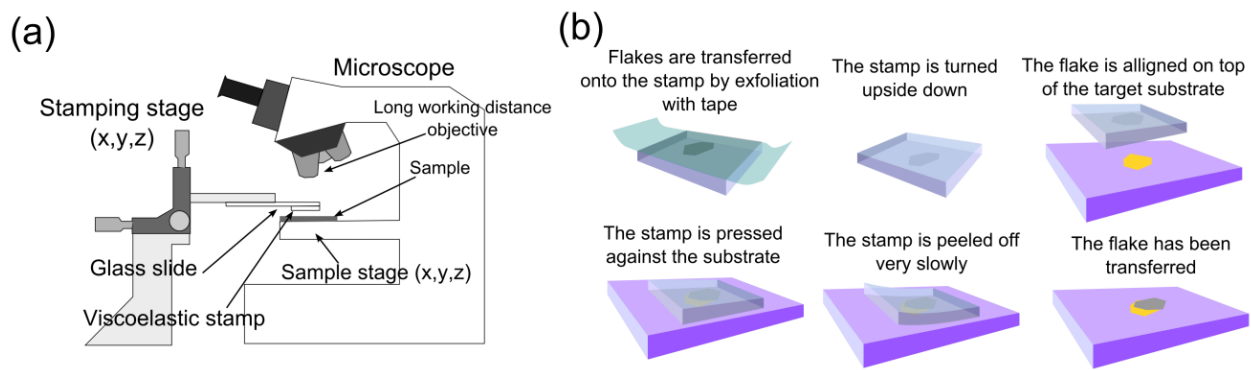

**Figure 1. Deterministic transfer setup and process.** (a) Schematic diagram of the experimental setup employed for the all-dry transfer process. (b) Diagram of the steps involved in the preparation of the viscoelastic stamp and the deterministic transfer of an atomically thin flake onto a user-defined location (for instance another atomically thin flake).

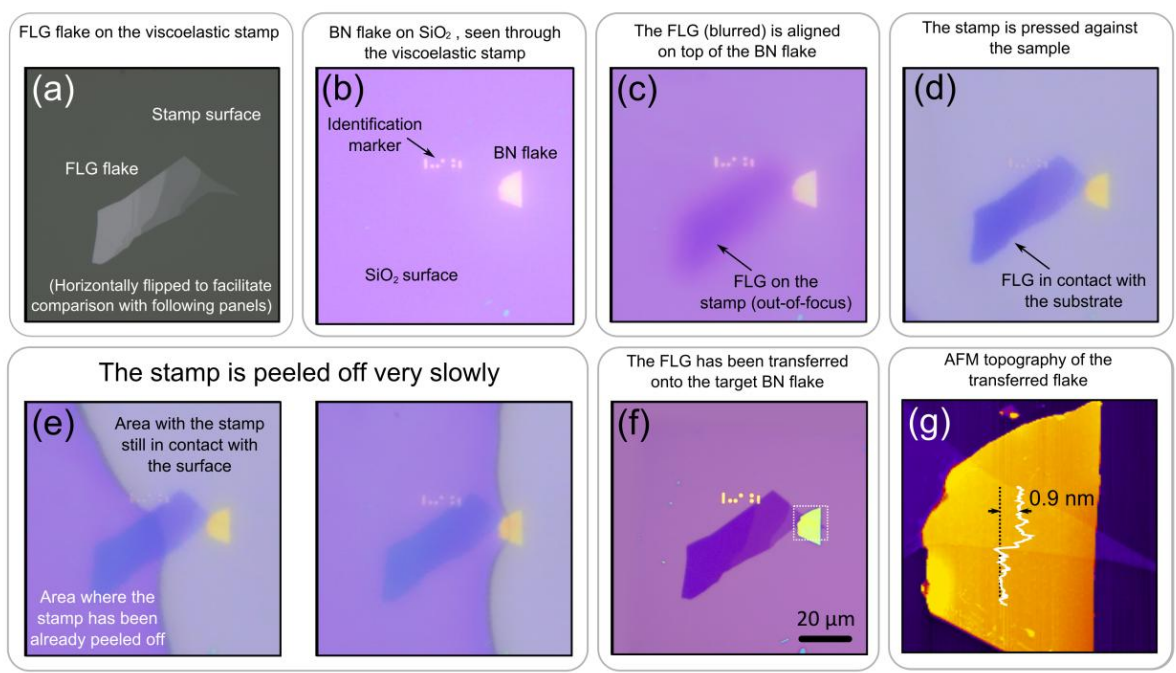

**Figure 2. Placing a few-layer graphene flake onto a hexagonal boron nitride flake.** (a) To (f) Optical micrographs acquired during the transfer process. (a) A few-layers graphene flake has been identified on the surface of the viscoelastic stamp. (b) A boron nitride flake has been selected on the acceptor surface. (c) The few-layers graphene flake is aligned on top of the boron nitride with a three axis micrometer positioner and looking through the transparent stamp during the handling. (d) Once the flake is aligned, the stamp is brought into contact with the sample. During this process the few-layer graphene flake becomes more and more focused. (e) The viscoelastic stamp is peeled off very slowly until the few-layer graphene flake is fully transferred. (f) Resulting heterostructure of few-layer graphene onto a boron nitride flake. (g) AFM topography of the transferred FLG flake (acquired on the area highlighted with a dashed square in (f)). A topographic line profile has been included as an inset.

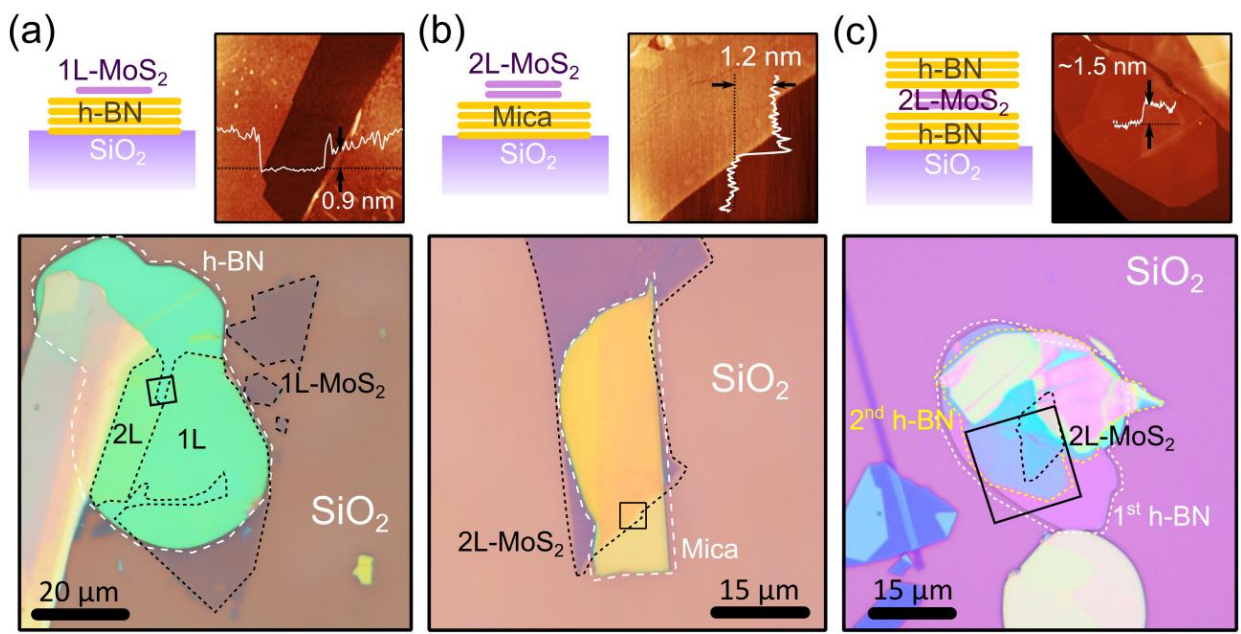

**Figure 3. Artificial heterostructures**. Optical and AFM images of some heterostructures fabricated by stacking different two-dimensional crystals (h-BN, $MoS_2$ and mica) with our all-dry transfer technique. A sketch of the fabricated heterostructure is shown above the optical images (at the left side). Atomic force microscopy images (acquired in the regions highlighted by a black square in the optical images) are also shown above the optical images (at the right side). (a) mono- and bilayer $MoS_2$ onto h-BN. (b) Bilayer $MoS_2$ onto a muscovite mica flake. (c) a bilayer $MoS_2$ flake 'sandwiched' between two h-BN flakes.

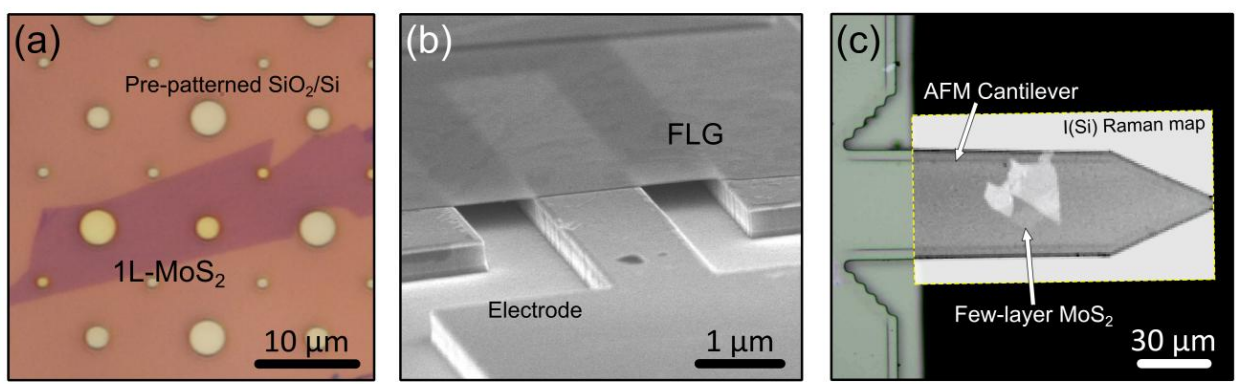

**Figure 4. Transfer of atomically-thin crystals onto different substrates**. (a) Single layer $MoS_2$ flake transferred onto a $SiO_2$/Si substrate pre-patterned with holes of 1, 2 and 3 μm in diameter. The $MoS_2$ flake covering the holes is freely suspended forming a micro drumhead. (b) Few-layer graphene (~ 10 layers) flake transferred onto a device, suspended between drain and source electrodes. (c) Few-layer $MoS_2$ flake (3 to 15 layers) transferred onto an freely overhanging AFM cantilever. A Raman map with the intensity of the silicon peak has been superimposed to the optical image of the cantilever to facilitate the identification of the transferred flake (region where the silicon peak has lower intensity due to the absorption of the $MoS_2$ flake) as it was almost invisible in the optical image.

REFERENCES

[1] Novoselov, K. S.;Geim, A. K.;Morozov, S.;Jiang, D.;Zhang, Y.;Dubonos, S.;Grigorieva, I.; Firsov, A. Electric field effect in atomically thin carbon films. *Science* **2004**, *306*, 666-669.
[2] Geim, A. K.; Novoselov, K. S. The rise of graphene. *Nature Materials* **2007**, *6*, 183-191.
[3] Novoselov, K.;Jiang, D.;Schedin, F.;Booth, T.;Khotkevich, V.;Morozov, S.; Geim, A. Two-dimensional atomic crystals. *Proceedings of the National Academy of Sciences of the United States of America* **2005**, *102*, 10451-10453.
[4] Novoselov, K. S.;Jiang, Z.;Zhang, Y.;Morozov, S.;Stormer, H.;Zeitler, U.;Maan, J.;Boebinger, G.;Kim, P.; Geim, A. Room-temperature quantum Hall effect in graphene. *Science* **2007**, *315*, 1379-1379.
[5] Blake, P.;Hill, E.;Neto, A. C.;Novoselov, K.;Jiang, D.;Yang, R.;Booth, T.; Geim, A. Making graphene visible. *Applied Physics Letters* **2007**, *91*, 063124.
[6] Roddaro, S.;Pingue, P.;Piazza, V.;Pellegrini, V.; Beltram, F. The optical visibility of graphene: Interference colors of ultrathin graphite on SiO2. *Nano Letters* **2007**, *7*, 2707-2710.
[7] Castellanos-Gomez, A.;Agraït, N.; Rubio-Bollinger, G. Optical identification of atomically thin dichalcogenide crystals. *Applied Physics Letters* **2010**, *96*, 213116.
[8] Novoselov, K.; Neto, A. C. Two-dimensional crystals-based heterostructures: materials with tailored properties. *Physica Scripta* **2012**, *2012*, 014006.
[9] Dean, C.;Young, A.;Wang, L.;Meric, I.;Lee, G.-H.;Watanabe, K.;Taniguchi, T.;Shepard, K.;Kim, P.; Hone, J. Graphene based heterostructures. *Solid State Communications* **2012**, *152*, 1275-1282.
[10] Grigorieva, A. G. I. Van der Waals heterostructures. *Nature* **2013**, *499*, 419-425.
[11] Britnell, L.;Ribeiro, R.;Eckmann, A.;Jalil, R.;Belle, B.;Mishchenko, A.;Kim, Y.-J.;Gorbachev, R.;Georgiou, T.; Morozov, S. Strong light-matter interactions in heterostructures of atomically thin films. *Science* **2013**, *340*, 1311-1314.
[12] Britnell, L.;Gorbachev, R.;Jalil, R.;Belle, B.;Schedin, F.;Mishchenko, A.;Georgiou, T.;Katsnelson, M.;Eaves, L.; Morozov, S. Field-effect tunneling transistor based on vertical graphene heterostructures. *Science* **2012**, *335*, 947-950.
[13] Yin, Z.;Li, H.;Li, H.;Jiang, L.;Shi, Y.;Sun, Y.;Lu, G.;Zhang, Q.;Chen, X.; Zhang, H. Single-layer MoS2 phototransistors. *ACS Nano* **2012**, *6*, 74-80.
[14] Georgiou, T.;Jalil, R.;Belle, B. D.;Britnell, L.;Gorbachev, R. V.;Morozov, S. V.;Kim, Y.-J.;Gholinia, A.;Haigh, S. J.; Makarovsky, O. Vertical field-effect transistor based on graphene-WS2 heterostructures for flexible and transparent electronics. *Nature Nanotechnology* **2012**, *8*, 100-103.
[15] Bonaccorso, F.;Lombardo, A.;Hasan, T.;Sun, Z.;Colombo, L.; Ferrari, A. C. Production and processing of graphene and 2d crystals. *Materials Today* **2012**, *15*, 564-589.
[16] Song, X.;Oksanen, M.;Sillanpää, M. A.;Craighead, H.;Parpia, J.; Hakonen, P. J. Stamp Transferred Suspended Graphene Mechanical Resonators for Radio Frequency Electrical Readout. *Nano Letters* **2011**, *12*, 198-202.
[17] Schneider, G. F.;Calado, V. E.;Zandbergen, H.;Vandersypen, L. M.; Dekker, C. Wedging transfer of nanostructures. *Nano Letters* **2010**, *10*, 1912-1916.
[18] Dean, C.;Young, A.;Meric, I.;Lee, C.;Wang, L.;Sorgenfrei, S.;Watanabe, K.;Taniguchi, T.;Kim, P.; Shepard, K. Boron nitride substrates for high-quality graphene electronics. *Nature Nanotechnology* **2010**, *5*, 722-726.
[19] Zomer, P.;Dash, S.;Tombros, N.; van Wees, B. A transfer technique for high mobility graphene

devices on commercially available hexagonal boron nitride. *Applied Physics Letters* **2011**, *99*, 232104.

[20] Ferrari, A. C.; Basko, D. M. Raman spectroscopy as a versatile tool for studying the properties of graphene. *Nature Nanotechnology* **2013**, *8*, 235-246.

[21] Lee, C.;Yan, H.;Brus, L. E.;Heinz, T. F.;Hone, J.; Ryu, S. Anomalous lattice vibrations of single-and few-layer MoS2. *ACS Nano* **2010**, *4*, 2695-2700.

[22] Meitl, M. A.;Zhu, Z.-T.;Kumar, V.;Lee, K. J.;Feng, X.;Huang, Y. Y.;Adesida, I.;Nuzzo, R. G.; Rogers, J. A. Transfer printing by kinetic control of adhesion to an elastomeric stamp. *Nature Materials* **2005**, *5*, 33-38.

[23] http://www.youtube.com/watch?v=cJ8NPE2Xcbg

[24] Goler, S.;Piazza, V.;Roddaro, S.;Pellegrini, V.;Beltram, F.; Pingue, P. Self-assembly and electron-beam-induced direct etching of suspended graphene nanostructures. *Journal of Applied Physics* **2011**, *110*, 064308.

[25] Buscema, M.;Steele, G. A.;van der Zant, H. S. J.; Castellanos-Gomez, A. Effect of the substrate on the Raman and Photoluminescence emission of single layer MoS2. *arXiv:1311.3869* **2013**.

[26] Castellanos-Gomez, A.;van Leeuwen, R.;Buscema, M.;van der Zant, H. S. J.;Steele, G. A.; Venstra, W. J. Single-Layer MoS2 Mechanical Resonators. *Advanced Materials* **2013**, *25*, 6719-6723

# Supporting Information:

# Deterministic transfer of two-dimensional materials by all-dry viscoelastic stamping

*Andres Castellanos-Gomez \*, Michele Buscema, Rianda Molenaar, Vibhor Singh, Laurens Janssen,*

*Herre S. J. van der Zant and Gary A. Steele*

Kavli Institute of Nanoscience, Delft University of Technology, Lorentzweg 1, 2628 CJ Delft (The Netherlands).

a.castellanosgomez@tudelft.nl

## Content

- **Materials and methods**
- **Step-by-step guide for transfer**
- **Real time video of the stamping process**
- **Sequential transfer: sandwiched structures**
- **More examples of heterostructures**
- **Strain induced bubbles**
- **Transfer onto $Si_3N_4$ membranes and holey carbon films**
- **Transfer onto pre-defined circuits**
    a. Measurement of two terminal resistance of graphene transferred on MoRe contacts
    b. Measurement of vertical tunnel junctions made by artificial stacking of 2D materials
- **Raman spectra of transfer samples**

1. **Materials and methods**

For our stamping method we employed commercially available elastomeric films as stamps. The films (Gel-Fim® WF ×4 6.0mil) were supplied by Gel-Pak. The Gel-Pak polymer is a polysiloxane based material similar to poly-dimethyl siloxane. In fact, we have characterized both Gel-Films® and home-made poly-dimethyl siloxane samples (made by curing Sylgard 184 elastomeric kit purchased from Dow Corning) by Raman spectroscopy finding that their spectra is very similar. This indicates that both polymers are very similar in structure. Moreover, we employed home-made poly-dimethyl siloxane stamps in an early stage finding similar results. Nevertheless, we found that the surface of the Gel-Films® is flatter than that of home-made poly-dimethyl siloxane sheets fabricated by casting Sylgard 184 elastomer onto a Petri dish.

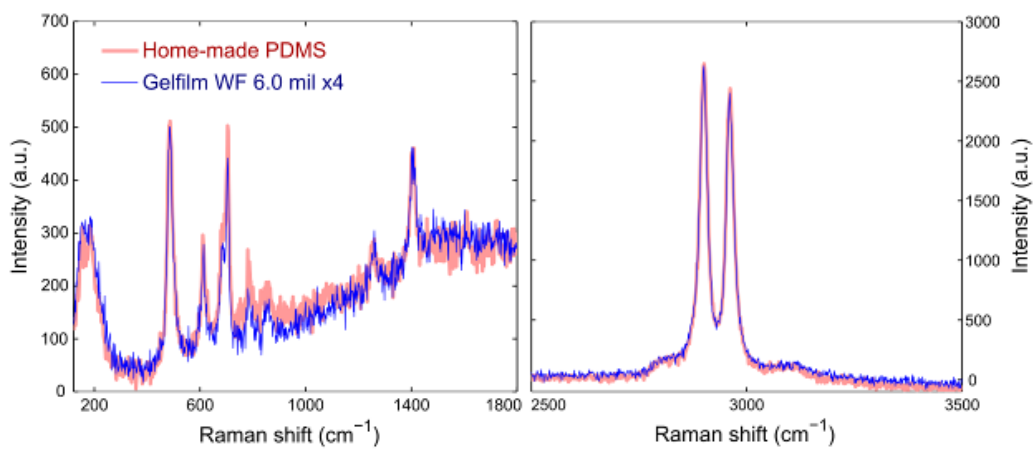

**Figure S1. Characterization of the elastomeric substrates.** Raman spectra acquired for an in-house fabricated poly-dimethyl siloxane substrate and for a commercially available Gel-Film® substrate.

The two-dimensional crystals are deposited onto the stamp surface by direct mechanical exfoliation of bulk layered materials (Graphite, $MoS_2$, h-BN and muscovite mica) with blue Nitto tape (Nitto Denko Co., SPV 224P). The employed bulk layered materials are: natural graphite flakes, natural $MoS_2$ (SPI Supplies, 429ML-AB), h-BN powder (Momentive, Polartherm grade PT110) and synthetic muscovite mica (grade V1). We located the atomically thin sheets under an optical microscope (Olympus BX 51 supplemented with an Canon EOS 600D digital camera) and estimated the number of layers by their opacity in transmission mode.

The topography of the transferred flakes has been characterized by atomic force microscopy and high-angle scanning electron microscopy. A Digital Instruments D3100 AFM (with standard cantilevers with spring constant of 40 N/m and tip curvature <10 nm) operated in the amplitude modulation mode has been used.

2. **Step-by-step guide for transfer**

    A. **Exfoliation**

We start by mechanically exfoliating the layered material:

- Preparation of the empty tape and the tape with material

- Bringing the two pieces of tape into contact

- Apply slight pressure uniformly with a cotton swab

- Mechanical cleaving the crystal by pulling the pieces of tape away from each other

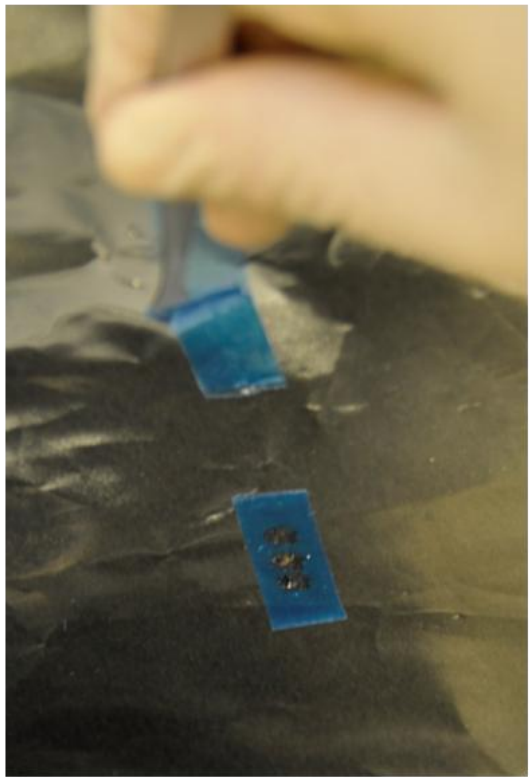
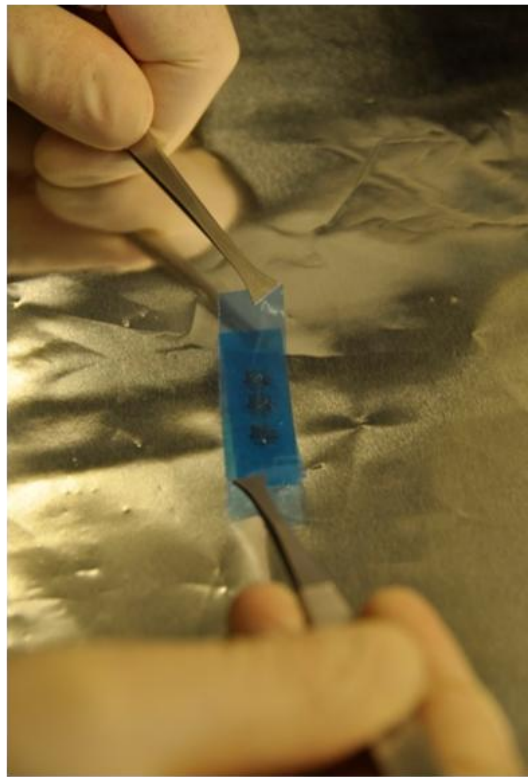
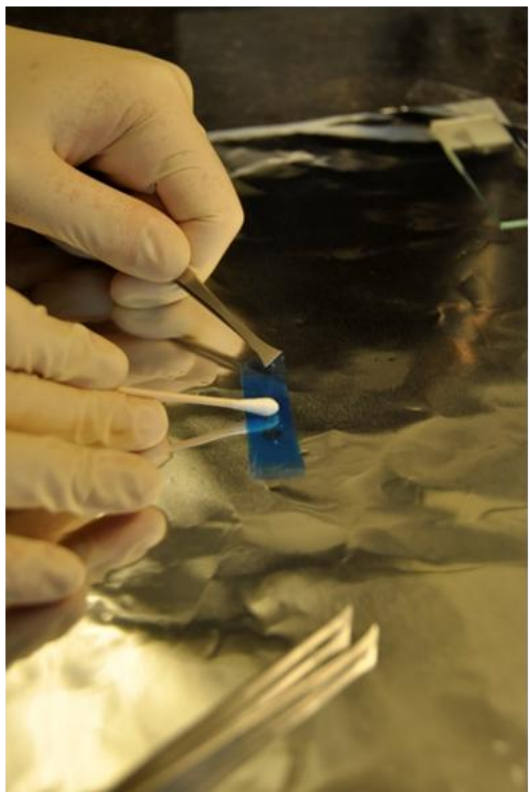
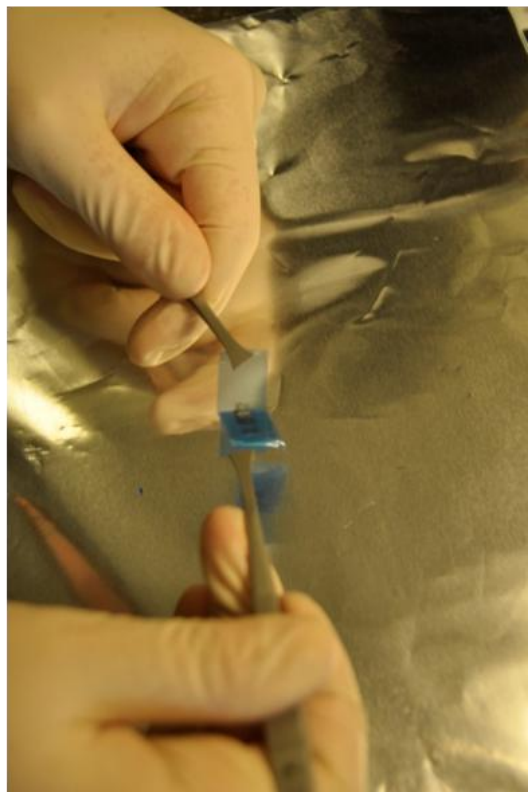

### B. Preparation of the viscoelastic stamp (1)

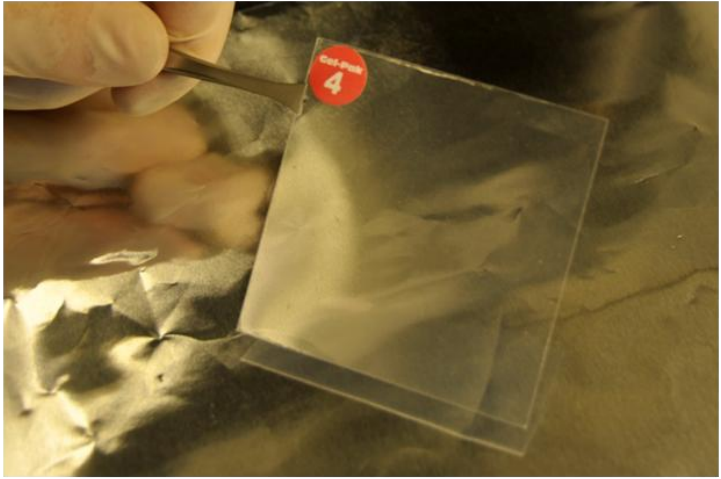

We start by taking a sheet of a commercial PDMS-based gel:
Supplier: Gel-Pak®
Part n: PF-3-X4
The gel is sandwiched between two protective polymeric layers

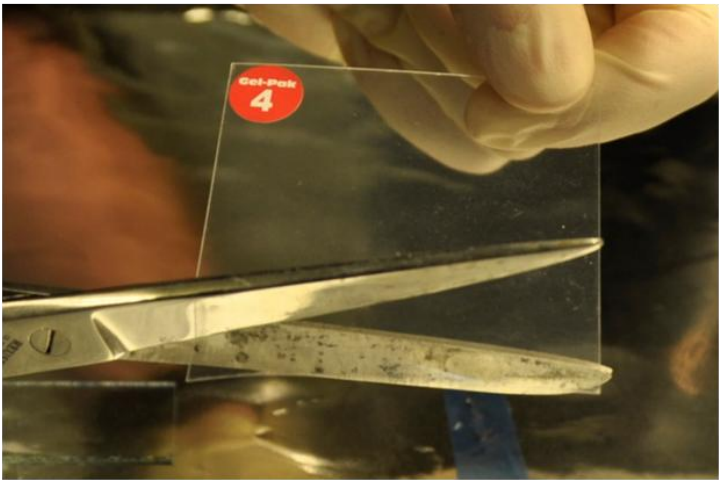

We proceed by cutting a piece out of the PDMS sheet

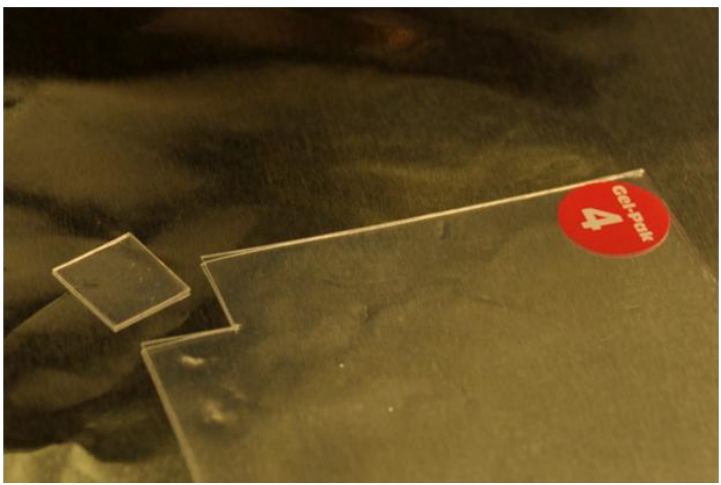

Detail of the cutout part

### C. Preparation of the viscoelastic stamp (2)

**1**

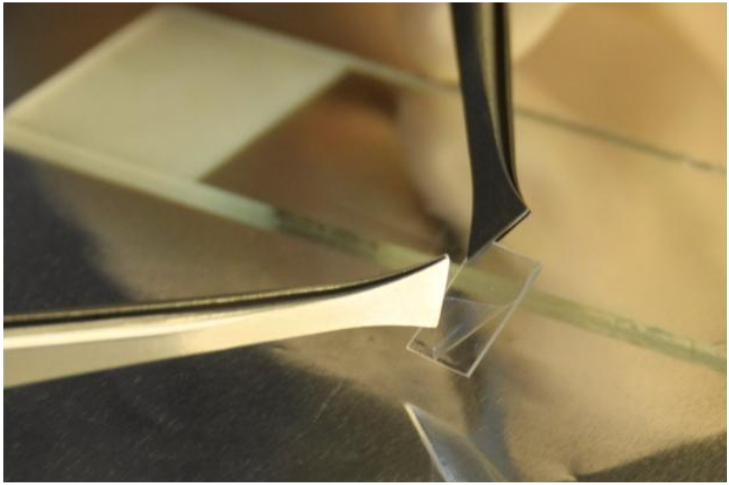

The first protective layer is removed

**2**

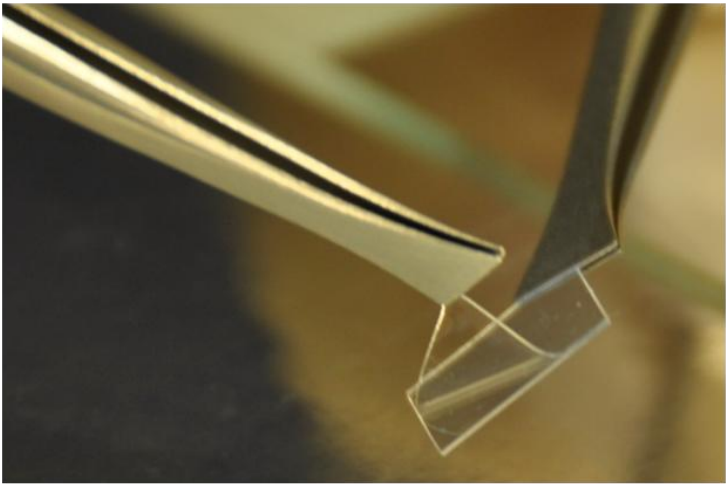

The PDMS gel is then slowly removed from the second protective layer

**3**

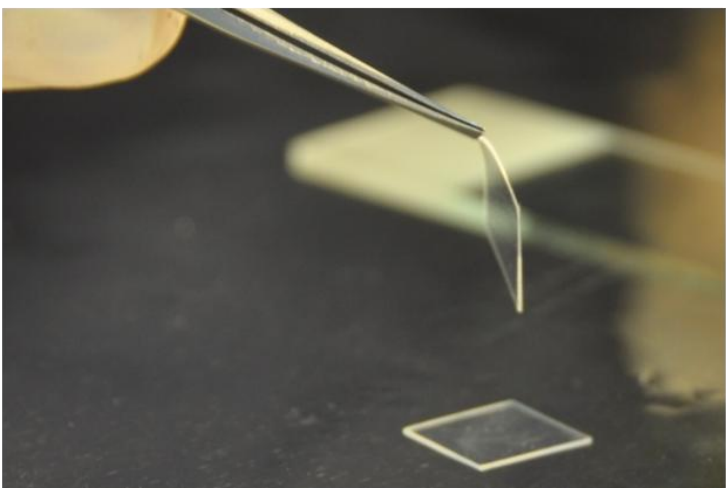

The PDMS gel is now freestanding and can be handled with tweezers

### D. Preparation of the viscoelastic stamp (3)

**1**

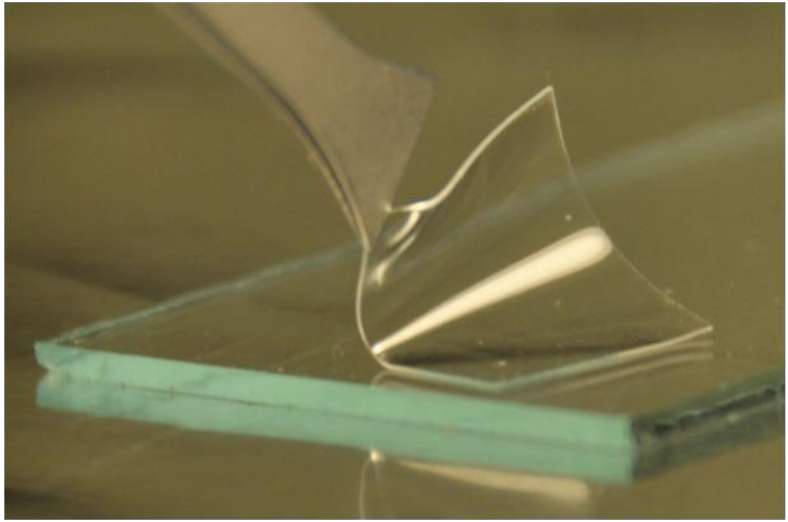

The PDMS gel is carefully brought in contact with the glass slide.

**2**

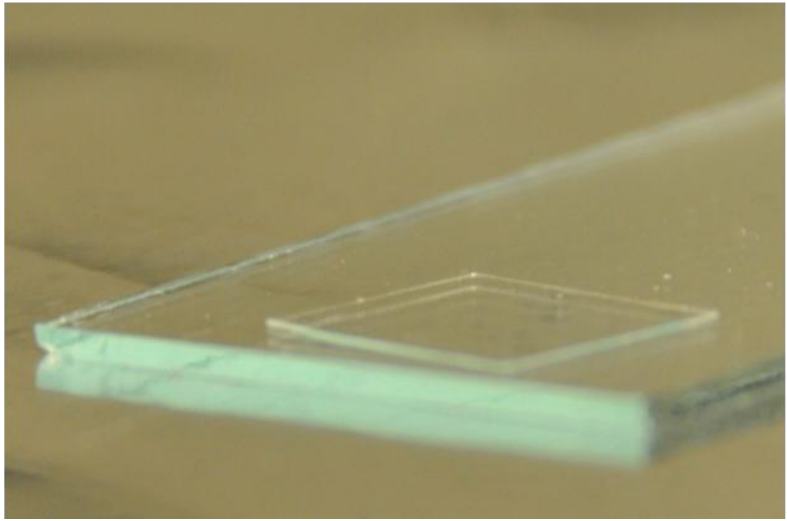

The PDMS gel adheres on the glass slide with no apparent wrinkles or bubbles.

### E. Transfer of flakes to the stamp

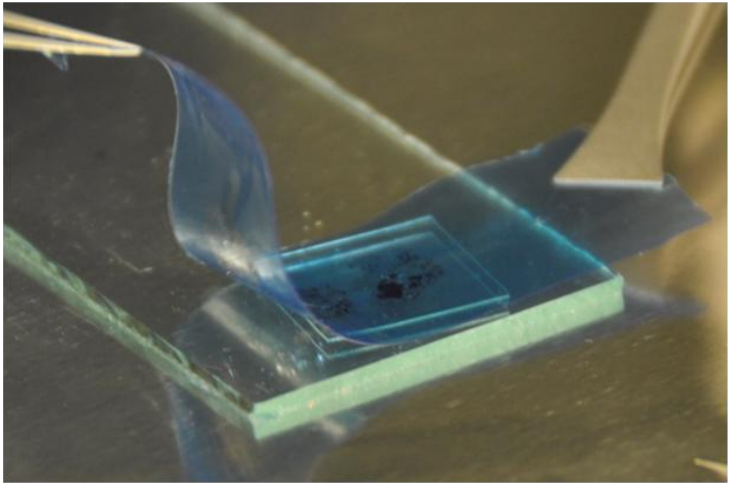

**1** The blue/scotch tape previously prepared is brought in contact with the PDMS stamp surface.

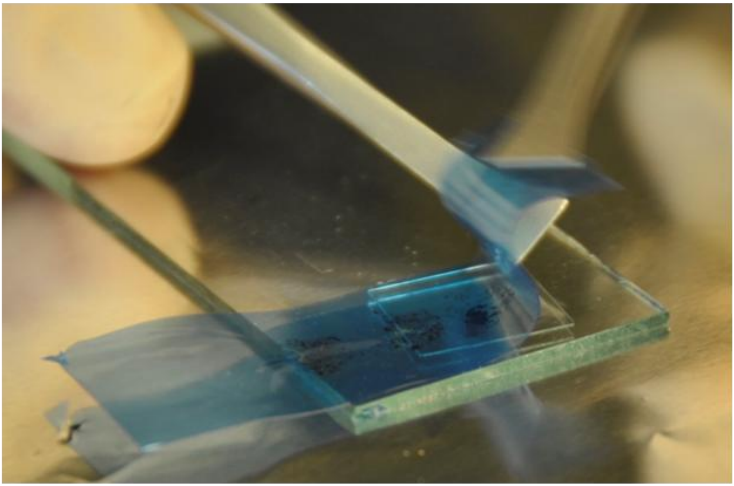

**2** While keeping the slide and the PDMS stamp into place, the blue tape with flakes is rapidly detached from the surface of the PDMS stamp.

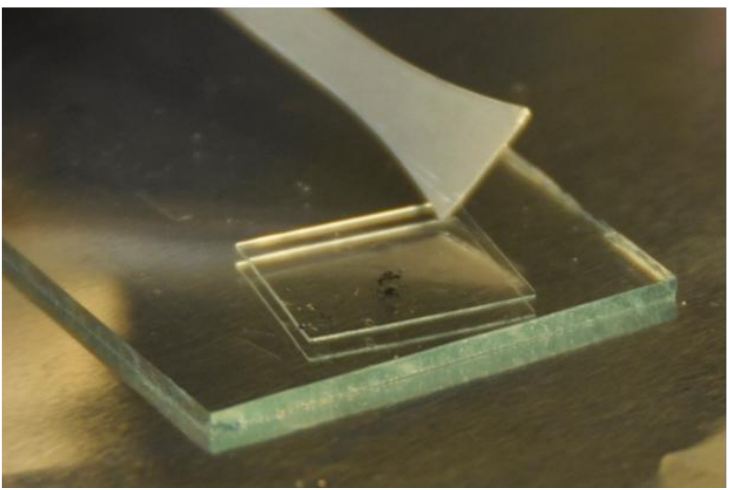

**3** After the removal of the tape, it is possible to see the transfer of some material. The stamp is now ready for optical inspection

### F. Optical identification

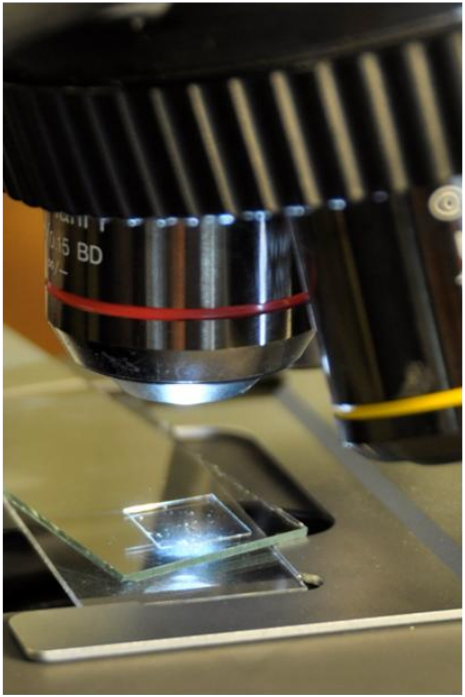

The stamp is transferred to an optical microscope that can be operated in both backscattering and transmission mode.

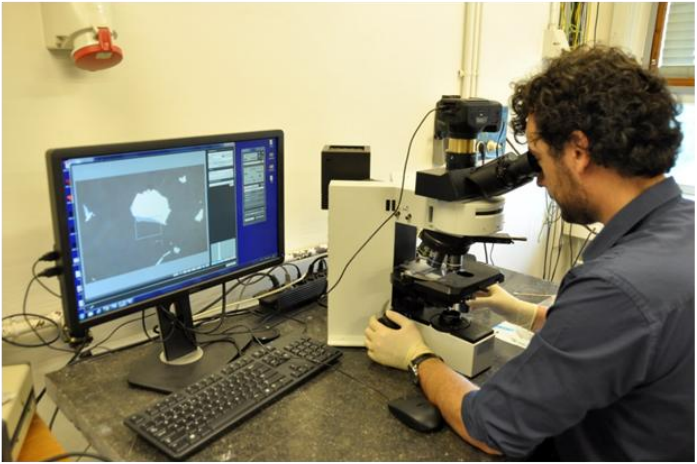

Typical appearance of a flake in backscattering (2) and transmission (3) optical microscopy

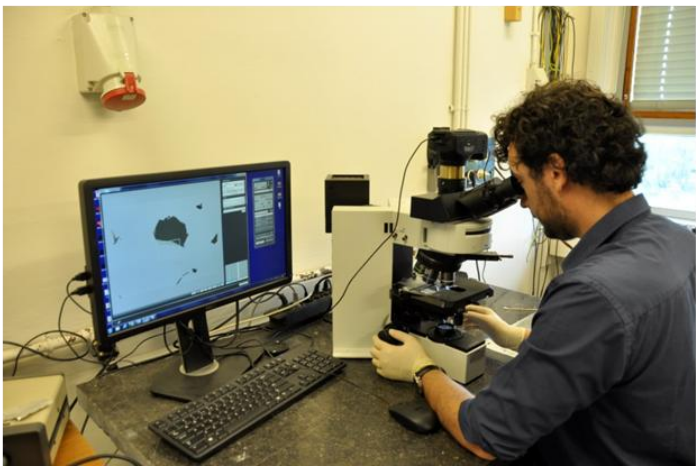

Once the desired flake is identified, the stamp can be used in the stamping setup

### G. Dedicated stamping setup

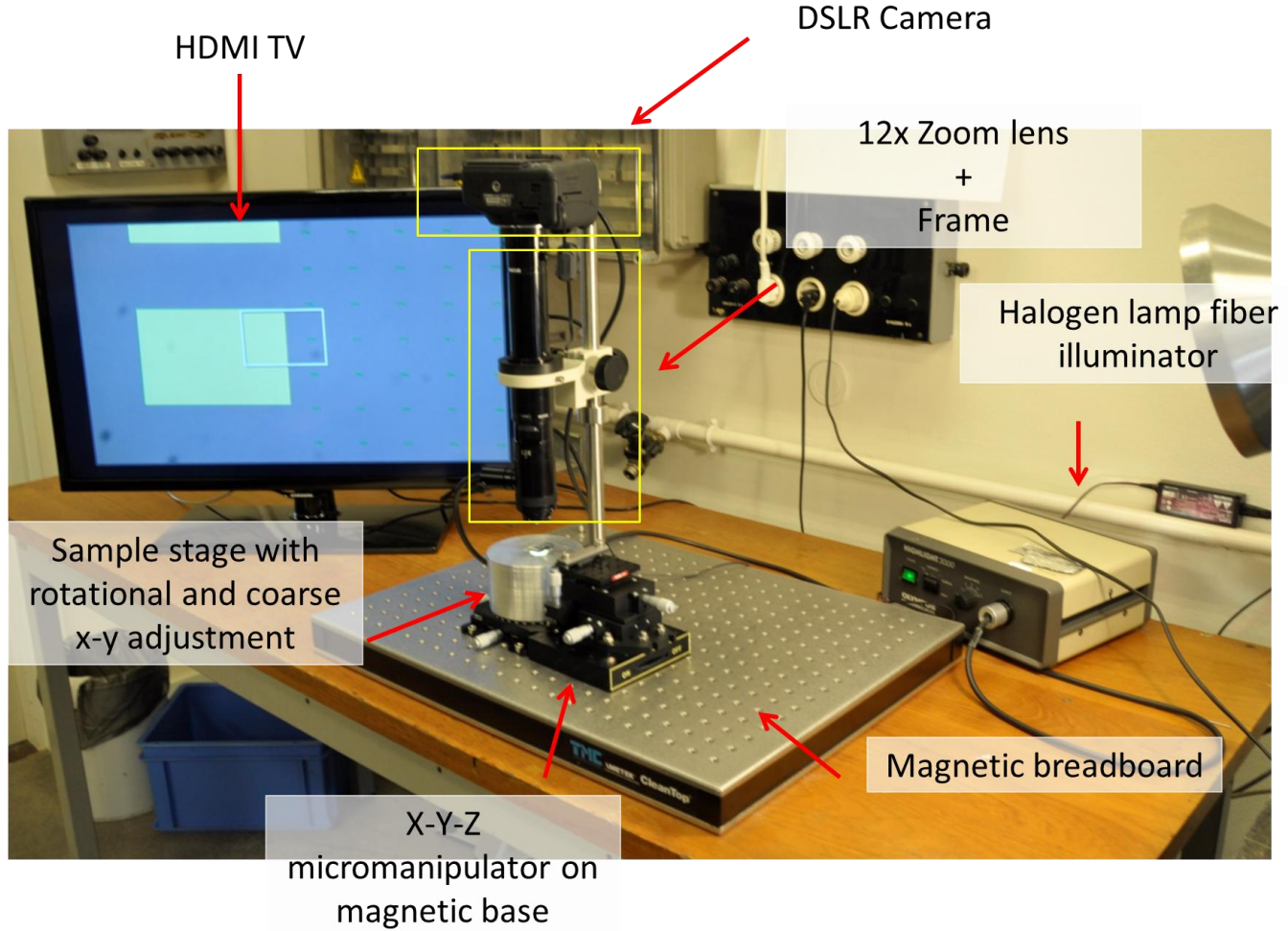

| Description | Part number | Price (€) | Supplier |
|---|---|---|---|
| Optical breadboard | #55-227 | 854.05 | Edmund Optics |
| Magnetic clamp | #62-272 | 369.55 | Edmund Optics |
| Table clamp | #54-262 | 75.05 | Edmund Optics |
| Steel bar | #39-353 | 52.25 | Edmund Optics |
| Rack and pinion | #03-609 | 237.5 | Edmund Optics |
| Extension Tube for Zoom lenses | MVL20FA | 542.88 | Thorlabs |
| Coaxially focusable Zoom lens | MVL12X3Z | 1887.9 | Thorlabs |
| Magnification lens | MVL12X20L | 281.88 | Thorlabs |
| XYZ micrometer stage | RB13M/M | 1152.72 | Thorlabs |
| Top plate for micrometer stage | RB13P1/M | 42.11 | Thorlabs |
| XY(theta) stage | XYR1/M | 494.16 | Thorlabs |
| Canon EOS600D | | 495 | |
| AC Charger (Canon EOS600D) | | 76.99 | |
| 16 GB SD card | | 29 | |
| Canon to F-mount addapter | | 8-20 | eBay |
| 32" TV Screen + HMDI adaptor | | 310 | |
| **Total** | | **6916** | |

### H. Stamping the desired flake (1)

**1**

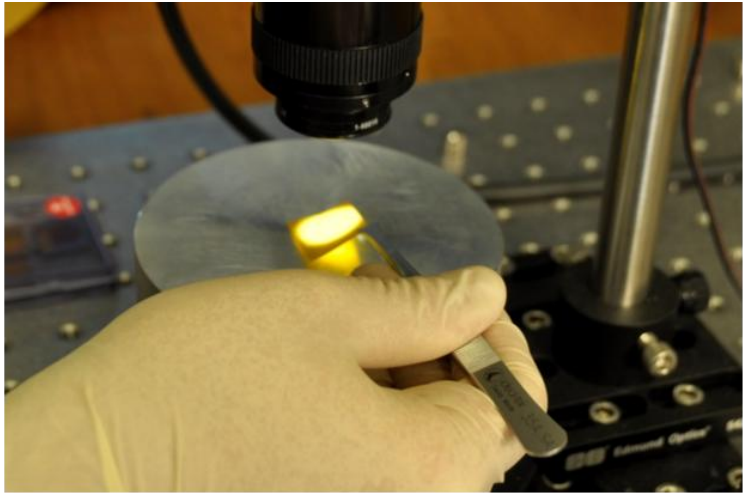

Double sided tape is placed on the sample stage

**2**

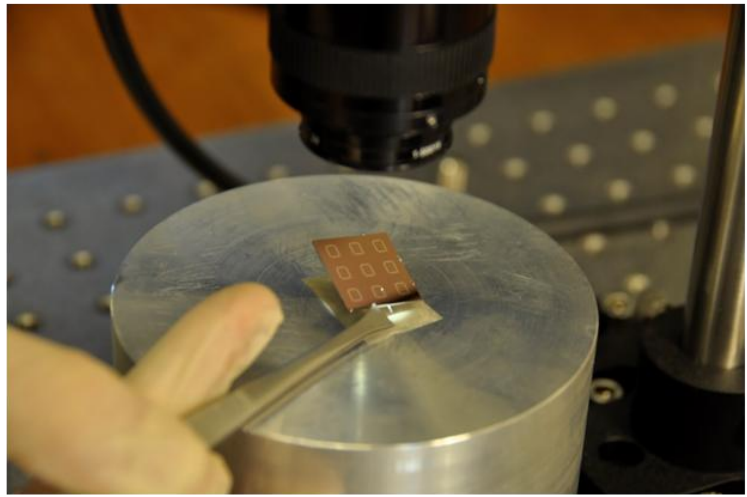

A patterned sample is placed on the sample stage

**3**

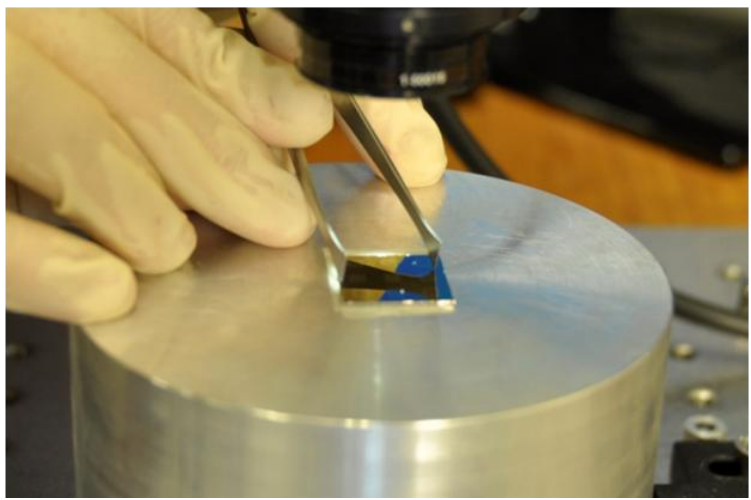

The sample is pressed on two corners to ensure flatness

I. **Stamping the desired flake (2)**

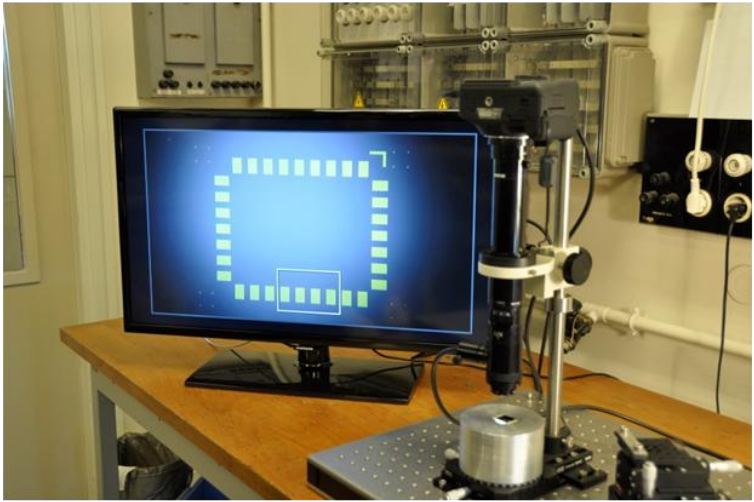

The pattern on the sample is imaged with the zoom lens

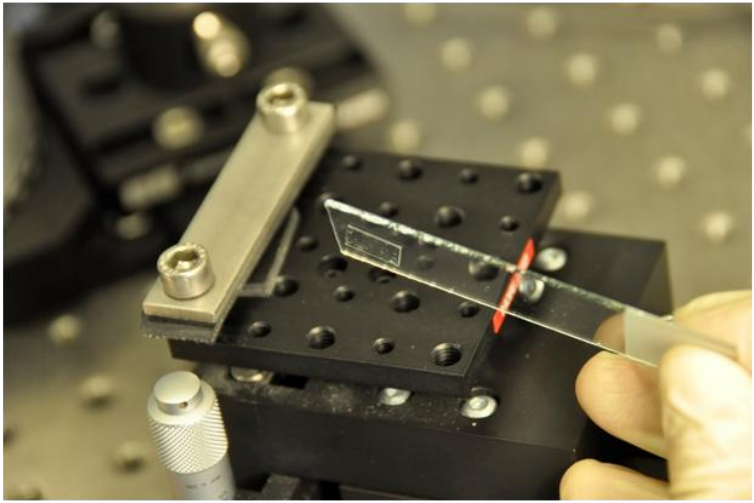

The stamp is rotated upside down

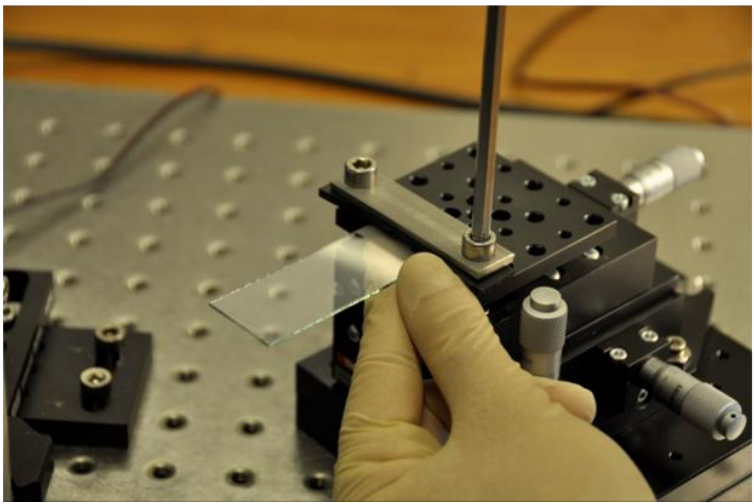

The stamp is mounted onto the micromanipulator

### J. Stamping the desired flake (3)

**1**

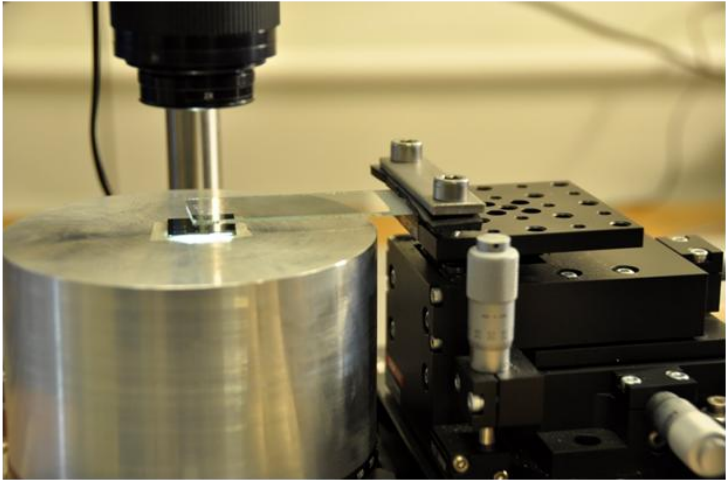

The stamp is coarsely positioned over the sample

**2**

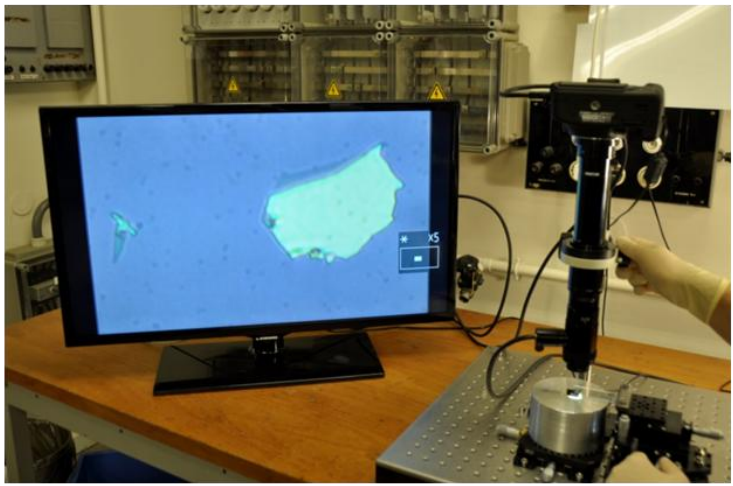

The desired flake is brought in the field of view of the zoom lens/camera assembly

**3**

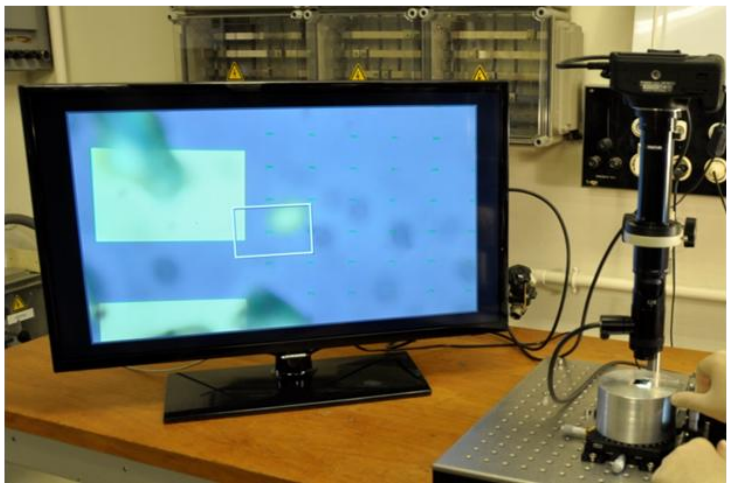

The desired pattern on the sample is brough into focus.

### K. Stamping the desired flake (4)

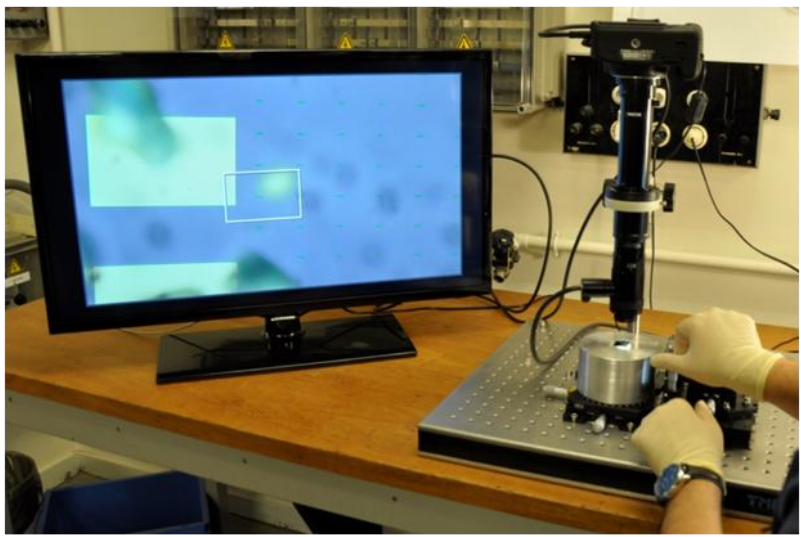

1. The stamp is then carefully lowered and adjusted in x-y with the micromanipulator so that the desired flake is aligned with the desired pattern

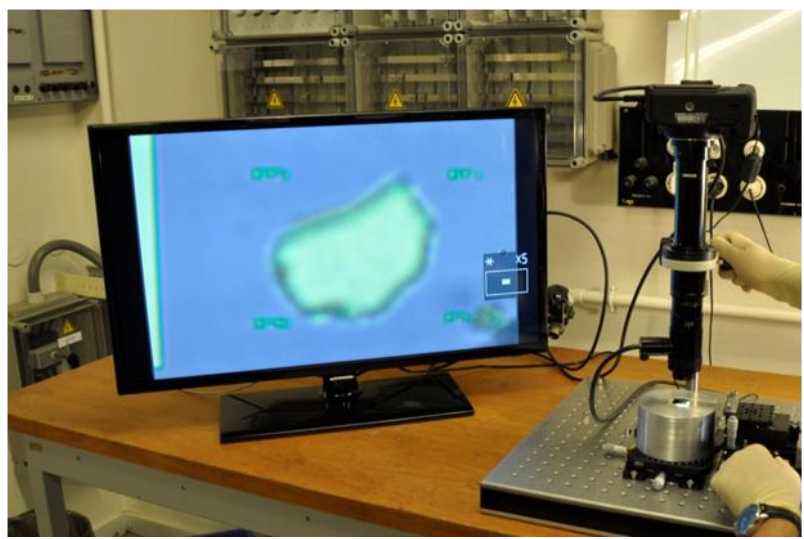

2. When the stamp is almost touching the surface, its position is finely adjusted with the help of the digital zoom of the DSLR camera.

Now the stamp is ready to be brought into contact with the sample surface. Please refer to the video also included in the Supplementary Information (see also following section).

3. **Real time video of the stamping process**

In the Supporting Information of this manuscript it has been also included a real time video of the stamping process. A single layer $MoS_2$ flake is transferred over a pre-patterned gold structure with a 8 µm hole. Figure S2 shows 12 frames of this video.

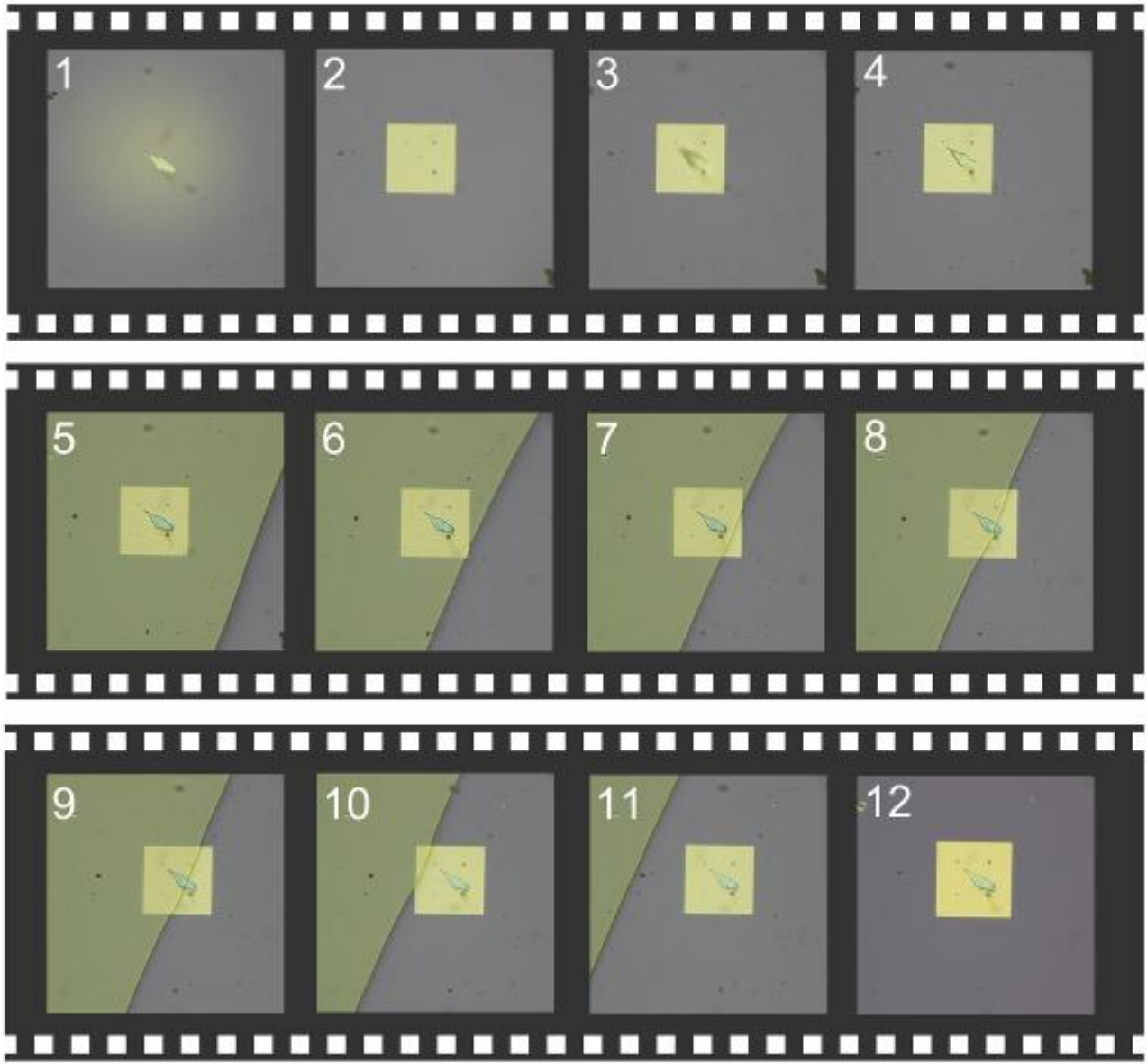

**Figure S2.** Frames of the real time video acquired during the stamping of a single-layer MoS$_2$ onto a pre-patterned substrate.

### 4. Sequential transfer: sandwiched structures

Figure S3 presents three optical images of the different steps in the fabrication of the 'sandwiched' MoS$_2$ structure shown in Figure 3d of the main text. A h-BN flake was first deposited onto a SiO$_2$(285 nm)/Si chip by mechanical exfoliation with Nitto tape (see Figure S3a). Then a bilayer MoS$_2$ flake was transferred onto the h-BN flake with the all-dry transfer technique (see Figure S3b). Finally, another h-BN flake was transferred onto the h-BN/MoS$_2$ heterostructure by another stamping step (see Figure S3c).

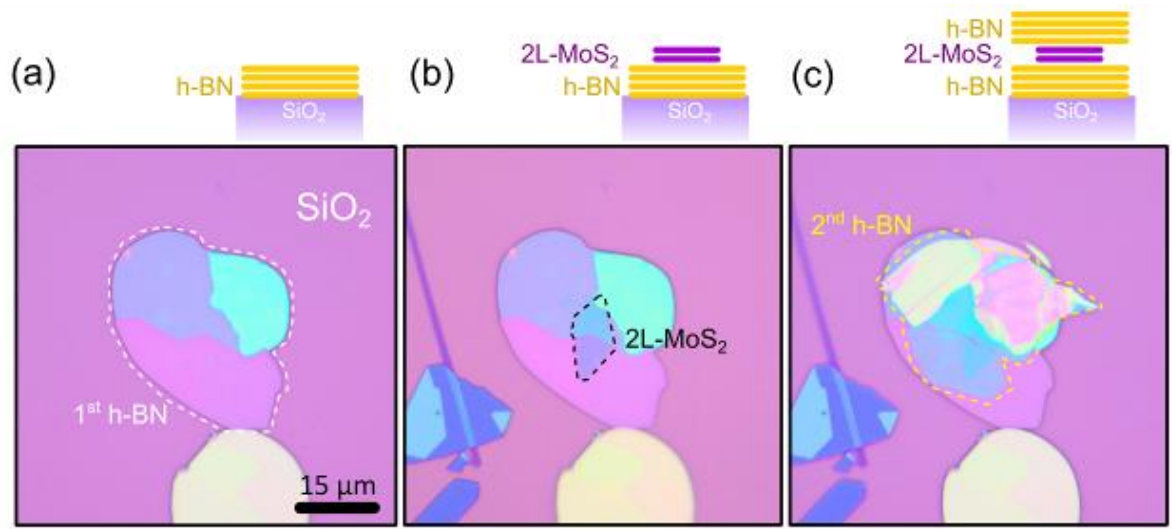

**Figure S3. Sequence to fabricate a sandwiched structure.** (a) Optical image of a h-BN flake transferred onto a $SiO_2$/Si chip by mechanical exfoliation. (b) Optical image of the same flake after transferring a bilayer $MoS_2$ flake. (c) Another h-BN flake is transferred onto the stack by employing another transfer step.

5. More examples of heterostructures

Figure S4 shows more examples of heterostructures, similar to those shown in Figure 3 of the main text.

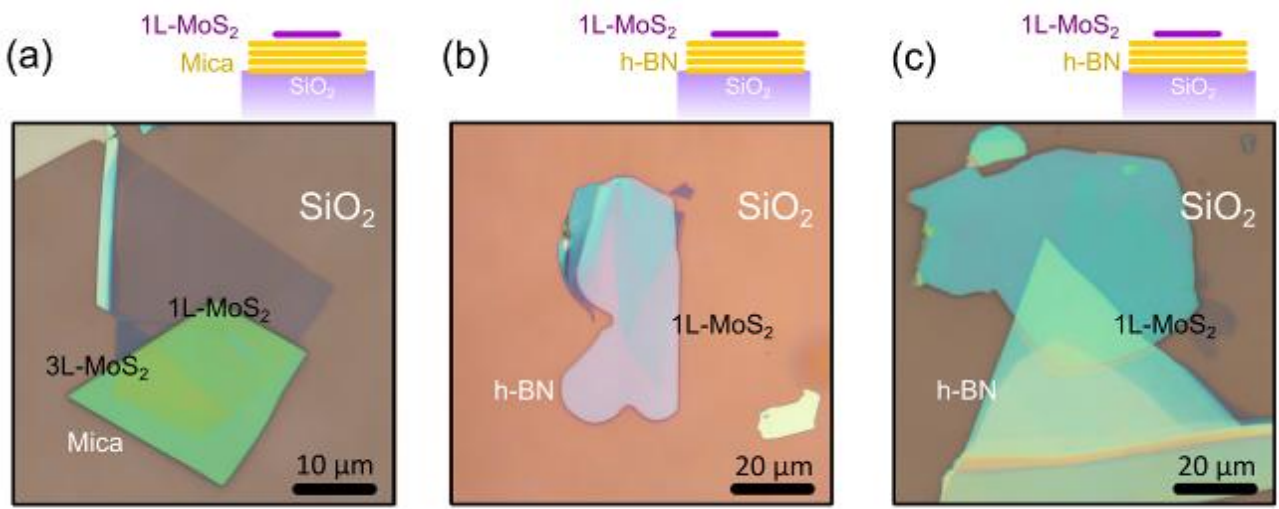

**Figure S4. Artificial stacking of 2D materials.** More examples of heterostructures fabricated with this technique. (a) Single-layer $MoS_2$ on mica. (b) and (c) Single-layer $MoS_2$ on h-BN.

6. **Strain induced bubbles**

During this work we have found that about 60%-70% of the fabricated heterostructures presented bubbles with diameters about 50 nm to 250 nm and height of 1 nm to 5 nm. These bubbles were present on the heterostructures fabricated by applying higher pressure to the stamp. We attribute the origin of these bubbles to the deformation of the viscoelastic stamp, which induces a large strain level on the transferred flake. Once the stamp is peeled off, this strain can be suddenly released forming these bubbles. A similar behaviour has been reported in Ref. [1] for graphene flakes fabricated by mechanical exfoliation with PDMS layers. Figure S5 shows an example of a single-layer $MoS_2$ flake with bubbles induced by the deformation of the stamp during the transfer.

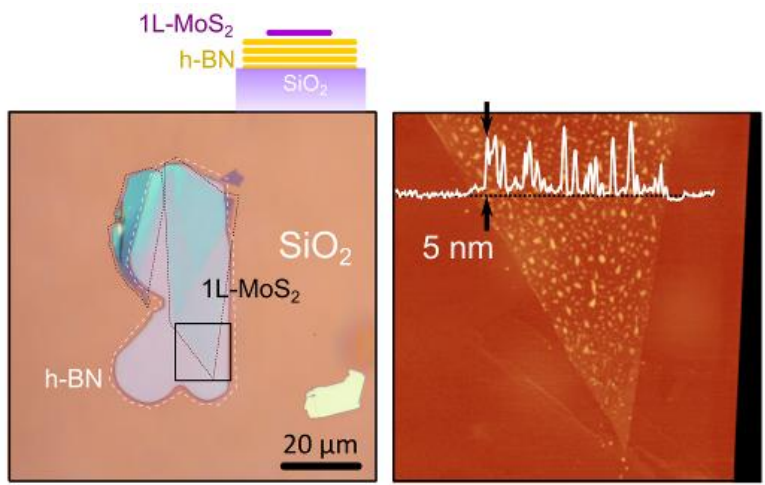

**Figure S5. Strain induced bubbles.** (Left) Optical image of a singl-layer $MoS_2$ device transferred onto a h-BN flake. (Right) AFM image of the region highlighted in (a) with a black square. The $MoS_2$ layer presents bubbles about 5 nm height and 100 nm wide due to an excessive pressure during the transfer.

7. **Transfer onto $Si_3N_4$ membranes and holey carbon films**

Figure 4c of the main text showed how 2D materials can be transferred onto an AFM cantilever without damaging the cantilever. Figure S6 shows how we also transferred two-dimensional materials onto silicon nitride membranes and holey carbon films, typically employed in transmission electron microscopy, without breaking them. Note that these substrates are rather fragile due to their reduced thickness.

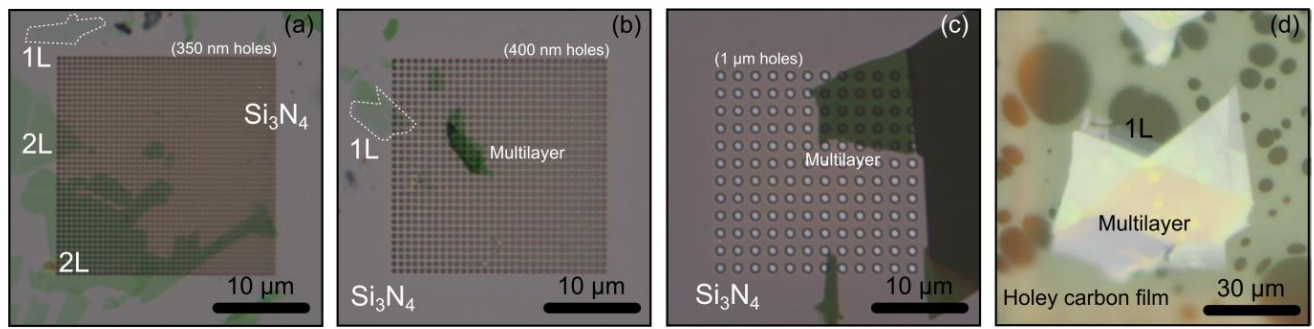

**Figure S6. MoS$_2$ transferred onto Si$_3$N$_4$ membranes and holey carbon films.** Ultrathin MoS$_2$ flakes have been gently transferred onto fragile substrates (silicon nitride Si$_3$N$_4$ membranes and holey carbon films) typically employed in transmission electron microscopy imaging. (a) To (c) Optical images (in transmission mode) of atomically thin MoS$_2$ flakes transferred onto 200 nm thick Si$_3$N$_4$ membrane with perforated holes of different diameters. (d) Optical images (in reflection mode) of an atomically thin MoS$_2$ flake transferred onto a holey carbon film.

## 8. Transfer onto pre-defined circuits

*Measurement of two terminal resistance of graphene transferred on MoRe contacts*

We have employed the dry-viscoelastic technique to transfer graphene on pre-patterned contacts. The contact pads were made of Molybdenum-Rhenium (60-40) alloy because its work function is similar to that of graphene. However, the graphene/MoRe interaction does not provide enough adhesion to perform the transfer. To modify the hydrophobicity of the metal surface, we coat the sample surface with a monolayer of HMDS (Bis(trimethylsilyl)amine). This surface treatment makes the graphene transfer very easy. Fig S7(a) shows the optical microscope image of a transferred few layer graphene flake on pre-patterned electrodes of MoRe (brighter area) fabricated on sapphire substrate (darker area).

The use of a self-assembled monolayer to increase the graphene adhesion has the drawback that the electrical contact between the graphene and the metal electrode is very poor and the transferred flakes are full of bubbles and wrinkles. The two terminal resistance is very high (~1 MOhm). To improve the contact between graphene and metal surface, we anneal the sample in a furnace at 150 ºC for 30 mins while flushing with hydrogen. The effect of annealing can be verified by optical microscopy: Figure S7(b) shows the optical microscope image of the same device after the annealing. Fig S7(c) shows the current voltage characteristic of the device at room temperature giving a resistance of ~400 Ohm which is a very low value if compared with other devices with similar geometry from the literature [2]. Table S1 below lists a comparison of two terminal resistance before and after annealing for all the devices studied.

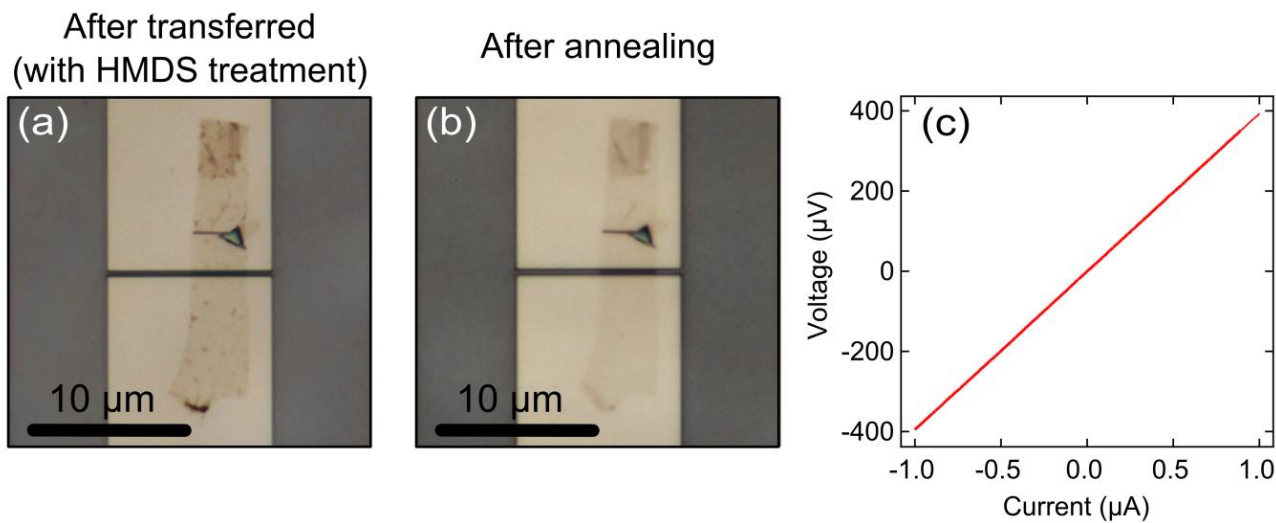

**Figure S7.** (a) Optical microscope image of two terminal graphene device before annealing. (b) After annealing to remove the HMDS and improve the contact. c) I-V characteristics taken at room temperature indicating two terminal resistance to be 400 Ohm.

**Table S1**

| Device | Resistance before annealing | Resistance after annealing | Distance between the electrode |
|---|---|---|---|
| Device 1 | 1 MOhm | 400 Ohm | 500 nm |
| Device 2 | 500 KOhm | 700 Ohm | 500 nm |
| Device 3 | 100 KOhm | 500 Ohm | 500 nm |

*Measurement of vertical tunnel junctions made by artificial stacking of 2D materials*

The electrical properties of transferred 2D materials have been further characterized by fabricating vertical tunnel junctions. A substrate with two pre-patterned metallic electrodes (5 nm Ti/ 50 Au) is employed. A thin $MoS_2$ flake (three layers thick) is transferred onto one of the electrodes. In a second transfer step, a 2.4 nm thick few-layer graphene (FLG) flake is transferred bridging the two electrodes. The FLG is carefully aligned to act as a top-electrode of a tunnel junction formed by the $Au/MoS_2/FLG$ stack. Figure S8a shows an optical image of the fabricated device. The inset in Figure S8b shows a schematic of the cross-section of the device.

The current vs. voltage characteristics of the device are measured in dark configuration (black line in Figure S8b). Then, the characteristics are measured again upon illumination with a green laser ($\lambda$ = 532 nm, 200 μm spot size and 750 μW). For large applied bias voltage, the device responds to the illumination with an increase of current with respect to the dark conditions. The maximum photoresponsivity of the device is 7.2 $AW^{-1}$ (measured at 0.4 V). This value is higher than the one of few-layer $MoS_2$ horizontal devices (0.6 $AW^{-1}$) [3] and only a factor of two lower than the photo-response reported for single-layer MoS$_2$ at similar power intensities (~18 AW$^{-1}$) [4]. Moreover, at zero bias voltage the MoS$_2$ vertical junction shows non-zero current due to photovoltaic effect which arises from the built-in potential due to the asymmetric Schottky barriers at the source and drain electrodes [5].

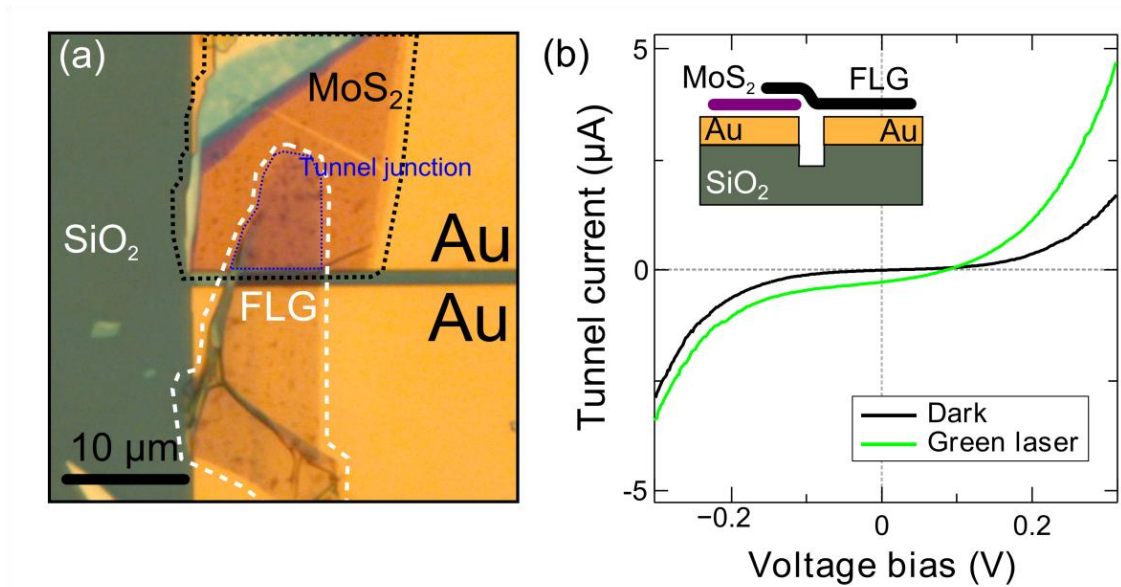

**Figure S8.** (a) Optical image of the vertical Au/MoS$_2$/FLG tunneling device. (b) Current vs. voltage characteristics of the device in dark (black line) and illuminated state (green line). A green laser (λ = 532 nm, 200 μm spot size and 750 μW) has been used to illuminate the device. (inset in b) Schematic cross-section of the fabricated vertical tunnel device.

9. **Raman spectra of transfer samples**

In this section, we present the Raman characterization of the MoS$_2$ flakes deposited via the all dry transfer method and compare it with the characterization of MoS$_2$ flakes on SiO$_2$ obtained via the standard micromechanical cleavage mechanism. Measurements are performed in a micro-Raman spectrometer (*Renishaw in via*) in backscattering configuration with 100x magnification objective (NA=0.95). Excitation is provided by an Argon laser (λ = 514 nm) with a typical power of ~ 200 μW. The resolution is in the order of 0.5 cm$^{-1}$.

.

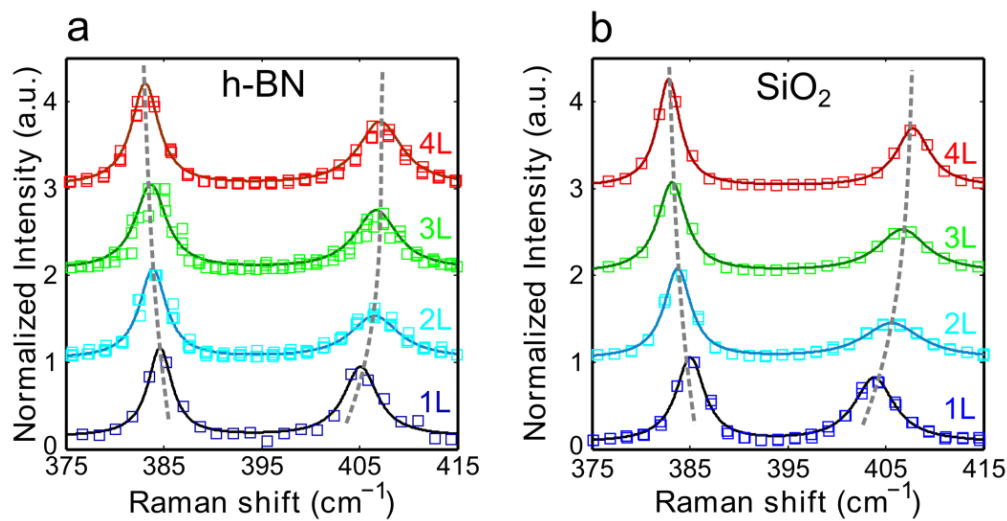

**Figure S9. Raman characterization of the transferred and directly exfoliated flakes.** (a) Normalized Raman intensity as a function of Raman shift of 1 layer (blue squares), 2 layers (light blue squares), 3 layers (green squares) and 4 layers (red squares) of $MoS_2$ transferred on h-BN by the all-dry transfer method. (b) Similar to panel (a) but measured on $MoS_2$ flakes directly exfoliated on $SiO_2$. In both panels the intensity is normalized to the intensity of the $E^1_{2g}$ Raman mode. The dashed lines are a guide-to-the eye for the position of the $E^1_{2g}$ and $A_{1g}$ $MoS_2$ Raman modes.

Figure S9 plots the Raman spectra for $MoS_2$ flakes transferred on h-BN (panel a) and directly exfoliated on $SiO_2$ (panel b). The measured points (open squares) are well fitted by Lorentzian functions (solid lines). The in-plane ($E^1_{2g}$ at ~ 380 $cm^{-1}$) and out-of-plane ($A_{1g}$ at ~ 405 $cm^{-1}$) Raman modes are clearly visible and their frequency changes with the number of layers. The difference between the $E^1_{2g}$ and $A_{1g}$ modes ($\Delta f$) is known to steadily increase with the number of layers and, therefore, it is a reliable quantity to count the number of layers of $MoS_2$ [6-10]. A more systematic characterization of the Raman spectra of $MoS_2$ samples transferred by the all-dry deterministic transfer method can be found in Ref. [11].

**Supporting information references**

[1]   Goler S, Piazza V, Roddaro S, Pellegrini V, Beltram F and Pingue P 2011 Self-assembly and electron-beam-induced direct etching of suspended graphene nanostructures *Journal of Applied Physics* **110** 064308--6

[2]   Burzurí E, Prins F and van der Zant H S 2012 Characterization of Nanometer-Spaced Few-Layer Graphene Electrodes *Graphene* **1** 26-9

[3]   Tsai D-S, Liu K-K, Lien D-H, Tsai M-L, Kang C-F, Lin C-A, Li L-J and He J-H 2013 Few Layer MoS2 with Broadband High Photogain and Fast Optical Switching for Use in Harsh Environments *ACS nano*

[4]   Lopez-Sanchez O, Lembke D, Kayci M, Radenovic A and Kis A 2013 Ultrasensitive photodetectors based on monolayer MoS2 *Nature Nanotechnology*

[5]   Fontana M, Deppe T, Boyd A K, Rinzan M, Liu A Y, Paranjape M and Barbara P 2013

       Electron-hole transport and photovoltaic effect in gated MoS2 Schottky junctions *Scientific reports* **3**

[6] Lee C, Yan H, Brus L E, Heinz T F, Hone J and Ryu S 2010 Anomalous Lattice Vibrations of Single- and Few-Layer MoS2 *ACS Nano* **4** 2695-700

[7] Cooper R C, Lee C, Marianetti C A, Wei X, Hone J and Kysar J W 2013 Nonlinear elastic behavior of two-dimensional molybdenum disulfide *Physical Review B* **87** 035423

[8] Mak K F, Lee C, Hone J, Shan J and Heinz T F 2010 Atomically Thin MoS_{2}: A New Direct-Gap Semiconductor *Physical Review Letters* **105** 136805

[9] Castellanos-Gomez A, Agrait N and Rubio-Bollinger G 2010 Optical identification of atomically thin dichalcogenide crystals *Applied Physics Letters* **96** 213116-3

[10] Splendiani A, Sun L, Zhang Y, Li T, Kim J, Chim C Y, Galli G and Wang F 2010 Emerging photoluminescence in monolayer MoS2 *Nano letters* **10** 1271-5

[11] Buscema M, Steele G A, van der Zant H S and Castellanos-Gomez A 2013 The effect of the substrate on the Raman and photoluminescence emission of single layer MoS2 *arXiv preprint arXiv:1311.3869*